\documentclass[conference]{IEEEtran}
\IEEEoverridecommandlockouts

\usepackage{cite}

\usepackage{float}
\usepackage[cmex10]{amsmath}
\usepackage{amsmath,amssymb,amsfonts}
\usepackage{mathtools}
\usepackage{mathdots}
\usepackage{mathrsfs}
\usepackage{algorithmic}
\usepackage{physics}
\usepackage{caption}
\usepackage{subcaption}
\usepackage{lipsum}

\usepackage{multirow}
\usepackage{multicol}

\usepackage{array}

\usepackage{graphicx}
\usepackage[font = small]{caption}
\usepackage{subcaption}

\usepackage{textcomp}
\usepackage{xcolor}
\def\BibTeX{{\rm B\kern-.05em{\sc i\kern-.025em b}\kern-.08em
    T\kern-.1667em\lower.7ex\hbox{E}\kern-.125emX}}

\usepackage[font = footnotesize]{caption}
\begin{document}

\title{HybridDeepRx: \\Deep Learning Receiver for High-EVM Signals}

\author{\IEEEauthorblockN{Jaakko Pihlajasalo$^1$,
Dani Korpi$^2$, Mikko Honkala$^2$, Janne M.~J. Huttunen$^2$, Taneli Riihonen$^1$,\\Jukka Talvitie$^1$, Alberto Brihuega$^1$, Mikko A. Uusitalo$^2$, and Mikko Valkama$^1$ \vspace{3pt} \\
$^1$Department of Electrical Engineering, Tampere University, Finland \\
$^2$Nokia Bell Labs, Espoo, Finland
}
}

\maketitle

\begin{abstract}
In this paper, we propose a machine learning (ML) based physical layer receiver solution for demodulating OFDM signals that are subject to a high level of nonlinear distortion. Specifically, a novel deep learning based convolutional neural network receiver is devised, containing layers in both time- and frequency domains, allowing to demodulate and decode the transmitted bits reliably despite the high error vector magnitude (EVM) in the transmit signal. Extensive set of numerical results is provided, in the context of 5G NR uplink incorporating also measured terminal power amplifier characteristics. The obtained results show that the proposed receiver system is able to clearly outperform classical linear receivers as well as existing ML receiver approaches, especially when the EVM is high in comparison with modulation order. The proposed ML receiver can thus facilitate pushing the terminal power amplifier (PA) systems deeper into saturation, and thereon improve the terminal power-efficiency, radiated power and network coverage.
\end{abstract}


\vspace{-1mm}
\section{Introduction} 
\vspace{-1mm}
Improving the network coverage and terminal power-efficiency are of fundamental importance in all mobile cellular systems. This is particularly so in wide-area macro deployments as well as in emerging millimeter-wave (mmWave) networks due to the challenges with propagation losses and trade-offs between hardware implementation costs, power consumption and transmit signal quality. Specifically, in the current 4G LTE/LTE-Advanced and 5G NR networks, the uplink coverage is primarily limited by the available user equipment (UE) transmit power while still meeting the unwanted emission and transmit signal inband quality requirements \cite{2019Toskala5G}.

Interestingly, while the feasible transmit power in below 3~GHz networks is commonly limited by the out-of-band (OOB) emission measures, the role of the passband error vector magnitude (EVM) is becoming more and more critical when the networks evolve towards the mmWave and later even the sub-THz bands. This is primarily because the nonlinear distortion is subject to beamforming \cite{8794583} as shown through concrete measurements, e.g., in \cite{9108534}. Inspired by this, in this article we develop a novel machine learning (ML) aided physical-layer receiver, referred to as HybridDeepRx, for efficiently demodulating orthogonal frequency-division multiplexing (OFDM) signals subject to substantial transmitter distortion. Specifically, the HybridDeepRx receiver is equipped with processing layers in both time- and frequency-domains, such that high-EVM signals can still be demodulated and detected efficiently. Extensive set of numerical results are also provided, in the context of 5G NR uplink, where measurement-based power amplifier (PA) models are deployed while experimenting with different levels of saturation and corresponding nonlinear distortion in the transmitter system, utilizing HybridDeepRx as the base station (BS) receiver. Based on the obtained results, the proposed hybrid ML receiver system clearly outperforms the classical linear minimum mean-squared error (LMMSE) receiver as well as earlier ML-based receivers. Finally, we note for clarity that there are many alternative technical approaches for coverage enhancements \cite{3gpp_TR38830}, while in this work we specifically focus on new deep learning based physical layer receiver technology.

{\em Notation:} Matrices are represented with boldface uppercase letters and they can consist of either real- or complex-valued elements, i.e., $\mathbf{X} \in \mathbb{F}^{N\times M}$, where $\mathbb{F}$ stands for either $\mathbb{R}$ or $\mathbb{C}$.

\section{State of the Art}
\label{sec:sota}

ML-aided radio reception has already been considered in several works, which have investigated implementing certain parts of the receiver chain with learned layers. For instance, channel estimation with neural networks has been studied in \cite{neumann18,he18}, while \cite{chang19} utilizes convolutional neural networks (CNNs) \cite{LeCun15} for equalization. \mbox{ML-based} demapping has been considered in \cite{shental2019}, where it was shown to achieve nearly the same accuracy as the optimal demapping rule, albeit with greatly reduced computational cost. Some works also propose augmenting the receiver processing flow with deep learning components \cite{gao18,He19a,Samuel19a} and show improved performance in comparison to conventional benchmark receivers.

A fully convolutional neural network based receiver, entitled DeepRx, was proposed in \cite{Honkala21,Korpi21}, and it was shown to achieve high performance especially under sparse pilot configurations. In addition to that, there are also other ML-based solutions for learning larger portions of the receiver, such as the work in \cite{ye18}, where channel estimation and signal detection are carried out jointly using a fully-connected neural network. There it is shown that the proposed ML-based receiver outperforms the conventional receiver when there are few channel estimation pilots or when the cyclic prefix is omitted. In addition, it is shown to be capable of handling rather well with clipping noise, a type of hard nonlinearity. The work in \cite{zhao2018}, on the other hand, applies CNNs to implement a receiver that extracts the bit estimates directly from a linear time-domain RX signal by learning the discrete Fourier transform (DFT). The prospect of learning the transmitter and receiver jointly has also been investigated by various works \cite{oshea17,dorner18,Aoudia21,Felix18}. Such schemes do not assume any prespecified modulation scheme or waveform, but instead learn everything from scratch. Such end-to-end learning has been shown to have potential to outperform traditional heuristic radio links, e.g., by learning a better constellation shape \cite{Aoudia21} or by learning to communicate under a nonlinear PA \cite{Felix18}.

\begin{figure*}[t!]
	\centering
	\includegraphics[width=0.8\textwidth,trim={0cm 9cm 7cm 0cm},clip]{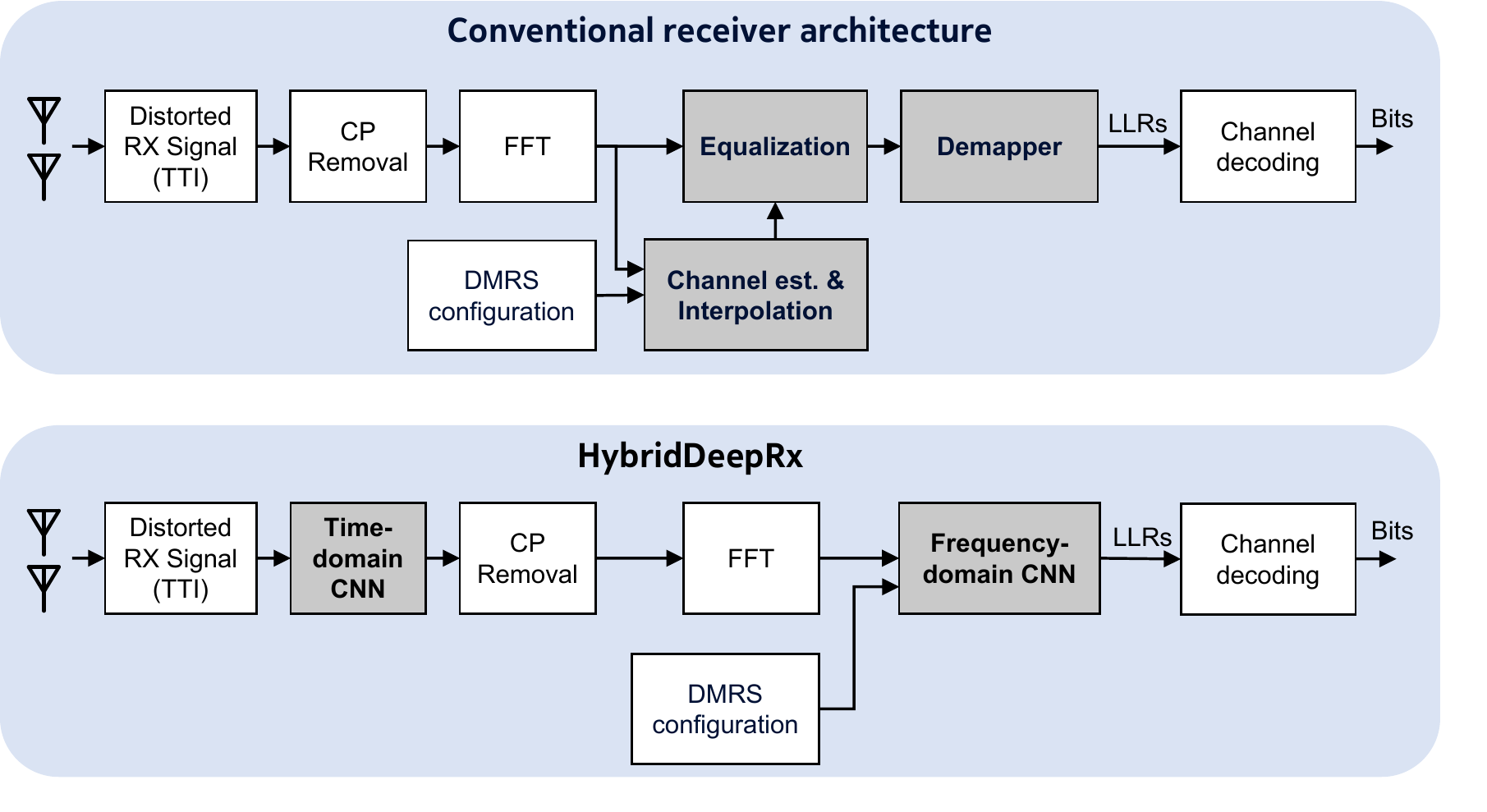}
	\vspace{-3mm}
	\caption{High-level depiction of a conventional OFDM receiver and of the proposed ML-based HybridDeepRx architecture.}
	\label{fig:general_blockdiag}
	\vspace{-2mm}
\end{figure*}

Despite the wide body of literature regarding ML-based radio receivers and the various demonstrations of their high performance, the effects of nonlinearities have been largely omitted in the analysis thus far. With the exception of the rather preliminary results in \cite{ye18} and the end-to-end learned system in \cite{Felix18}, there have been no tailor-made ML-based receivers for handling nonlinearly distorted RX signals, particularly when the level of distortion is substantially higher than what is allowed by the current 5G NR EVM specifications \cite{3gpp_TS38101-2}. In this paper, we fulfill this gap and show a specific CNN-based receiver architecture that is capable of accurate signal detection even under heavy PA-induced nonlinear distortion.

\section{System Model}

Figure~\ref{fig:general_blockdiag} depicts the general framework of the considered receiver architecture. The upper part of Fig.~\ref{fig:general_blockdiag} illustrates a conventional OFDM receiver, for reference, while the lower part shows the proposed receiver system with partially learned components. Let us first describe the basic signal model and also the conventional receiver processing. Using baseband-equivalent modeling, the received nonlinearly distorted time-domain signal can be expressed as
\begin{align}
    y(n) = h(n) * \phi \big(x(n)\big) + w(n),
\end{align}
where $h(n)$ denotes the multipath channel response, $*$ is the convolution operation, $\phi \left(\cdot\right)$ is the nonlinear response of the transmitter PA, $x(n)$ is the undistorted transmit waveform, and $w(n)$ is the noise-plus-interference signal. Considering the signal during a single transmission time interval (TTI), the received time-domain signal can be denoted by a matrix $\mathbf{Y}_t \in \mathbb{C}^{(N_{CP}+N)\times N_\mathrm{symb}}$, where $N_{CP}$ is the maximum cyclic prefix (CP) length within the TTI, $N$ is the FFT size and $N_\mathrm{symb}$ is the number of OFDM symbols. That is, the elements of $\mathbf{Y}_t$ consist simply of the received signal samples, ordered based on their corresponding OFDM symbols. In case the symbols have different CP lengths, zero-padding is used to align the total symbol lengths to $N_{CP}+N$.

Having first removed the CP, the signal is converted to its frequency-domain representation with a fast Fourier transform (FFT), after which it can be expressed as follows:
\begin{align}
\mathbf{Y}_f = \mathbf{H} \odot \textbf{X} + \mathbf{N}
  \label{eq:rx_signal}, 
\end{align}
where $\mathbf{Y}_f \in \mathbb{C}^{N_D \times N_\mathrm{symb}}$ and $\textbf{X} \in \mathbb{C}^{N_D \times N_\mathrm{symb}}$ are the received and transmitted symbols, respectively, $\mathbf{H} \in \mathbb{C}^{N_D \times N_\mathrm{symb}}$ is the frequency-domain channel matrix, $\odot$ denotes element wise multiplication, $\mathbf{N} \in \mathbb{C}^{N_D \times N_\mathrm{symb}}$ is the noise-plus-interference signal, and $N_D$ denotes the number of data-carrying subcarriers. The noise-plus-interference term incorporates also the effects of nonlinear distortion not captured by the linear part of the signal model.

In a conventional receiver, the demodulation reference signals (DMRSs) are extracted from $\mathbf{Y}_f$ for channel estimation, as illustrated in the upper part of Fig.~\ref{fig:general_blockdiag}, after which the signal is equalized and the soft bits are extracted. In this work, we consider the widely-used LMMSE receiver as the baseline or reference. For a description of such a receiver, see, e.g., \cite{Honkala21}. As a final outcome, the receiver will provide the so-called log-likelihood ratios (LLRs) for each data-carrying resource element (RE).

\vspace{-0.2mm}
\section{Proposed HybridDeepRx Receiver}
\label{sec:comp}
\begin{figure}[t!]
	\centering
	\includegraphics[width=0.75\columnwidth,trim={0cm 8cm 25.3cm 0cm},clip]{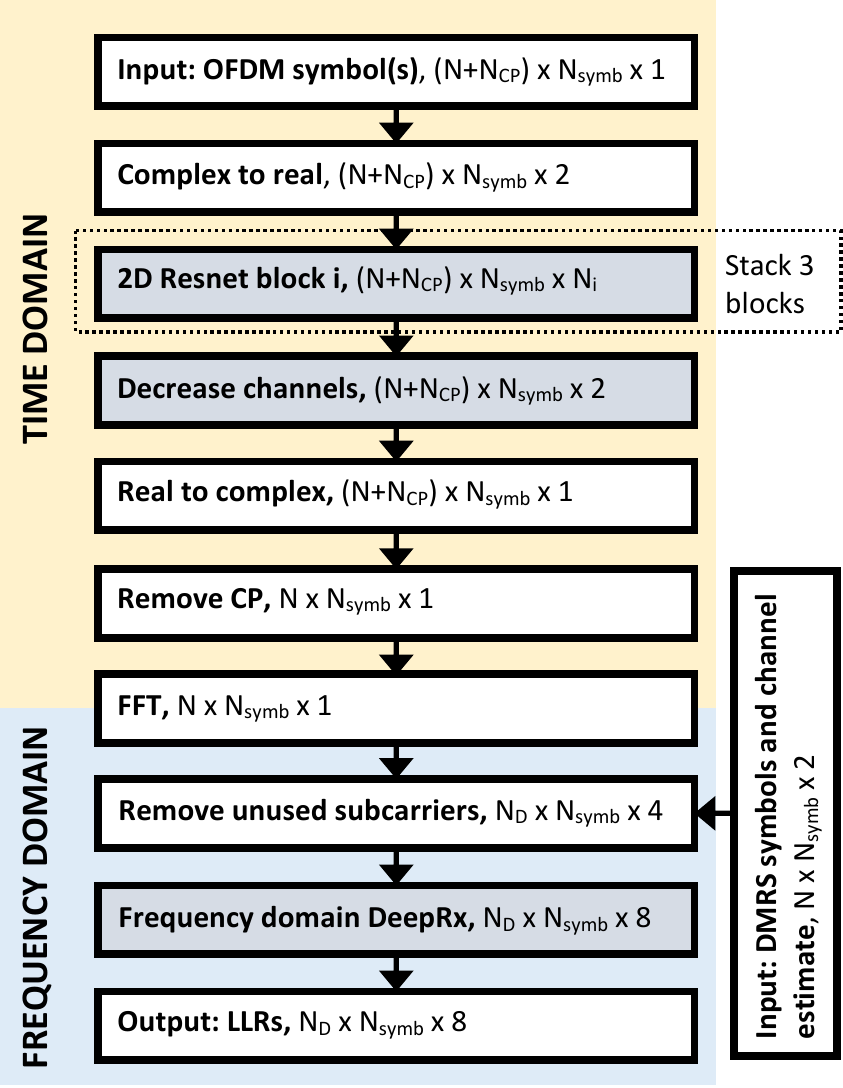}
	\caption{A detailed illustration of the HybridDeepRx receiver, where the dimensions within each block correspond to its output. The gray blocks represent the components of the learned architecture.}
	\label{fig:detailed_blockdiag}
\end{figure}
\vspace{-0.4mm}
The goal of the proposed neural network (NN)-based HybridDeepRx is to detect the raw bits from the nonlinearly distorted RX signals collected during a TTI, represented by the matrix $\mathbf{Y}_t$. A high-level depiction of the receiver architecture is shown in the lower part of Fig.~\ref{fig:general_blockdiag}. As the nonlinear distortion caused by the PA is a time domain phenomenon, we believe that neural network processing with time domain inputs is an efficient method for learning to detect such distorted signals. Moreover, since the target of the proposed receiver is to detect the transmitted data, which is obviously modulated in the frequency domain, the receiver utilizes NN-based processing in both time and frequency domains. This is achieved by including the FFT in the network as an untrainable layer, by which the complete receiver can be trained jointly.

A more detailed depiction of the NN-based receiver is shown in Fig.~\ref{fig:detailed_blockdiag}. The trainable parts of the receiver are indicated with gray blocks, and they follow a residual network (ResNet) structure \cite{he2016identity}. The part of the NN in-between the time- and frequency-domain ResNets consists of the CP removal and FFT, and they are performed as in a regular receiver. In the continuation, we shall refer to the two trainable parts of the network as pre-FFT and post-FFT networks.

The pre-FFT network takes the received time domain RX signals over one TTI as an input. Since the signals are complex valued, we concatenate the real and imaginary parts of the input along the third dimension. To take the varied CP lengths of the 5G specification \cite{3GPP_TR38211} into account, the OFDM signals with the shorter CP are padded with zeroes, to match the length of the longer CP. Thus the input to the pre-FFT network will be a real valued array $\mathbf{Z}_\mathrm{pre} \in \mathbb{R}^{(N_\mathrm{CP}+N)\times N_\mathrm{symb} \times 2}$, where $N_\mathrm{CP}$ again refers to the longest CP length. The pre-FFT network is built with 2D convolutional layers with residual connections using pre-activation ResNet \cite{he2016identity} blocks without batch normalisation layers. The network has three ResNet blocks with 64, 128 and 256 convolutional filters, respectively, and the size of the filters is $3\times3$. The dimensions of the outputs are kept the same size as those of the inputs. In Fig.~\ref{fig:detailed_blockdiag} the number of filters for ResNet block $i$ is denoted by $N_i$ (i.e., $N_1 = 64$, $N_2 = 128$, and $N_3 = 256$). The last layer of the pre-FFT network is a convolutional layer with two filters of size $1\times1$ and no activations. The output of the pre-FFT network is thus the same size as the input $\mathbf{Z}_\mathrm{pre}$, which represents the concatenated real and imaginary parts of the RX signal.

The output of the pre-FFT network is then fed to the FFT layers of the network, which represent the CP removal and FFT process of a regular receiver. First, the real and imaginary parts are combined to a complex-valued signal, after which the CP is removed. After the FFT, the unused subcarriers are removed, and the real and imaginary parts of the output are again concatenated and fed into the post-FFT network.

The post-FFT network takes in the output of the FFT layers and the raw least squares DMRS channel estimates, whose real and imaginary values are concatenated along the channel dimension (for the data-carrying REs, the raw channel estimate array contains zeros). Hence, the input to the post-FFT network is a real valued array $\mathbf{Z}_\mathrm{post} \in \mathbb{R}^{N_D\times N_\mathrm{symb} \times 4}$. This post-FFT network follows the DeepRx architecture presented in \cite{Honkala21}, which is also built with 2D preactivation ResNet blocks. The output is a real valued array $\mathbf{L} \in \mathbb{R}^{N_D \times N_\mathrm{symb} \times N_B}$ consisting of the detected LLRs, where $N_B$ is the maximum number of bits per symbol. In this work we have set $N_B = 8$ similar to \cite{Honkala21}.

\subsection*{Training Procedure}

The training is performed using the binary cross entropy (CE) as the loss function. Although the actual output of the post-FFT network consists of the LLRs, the training of the neural network is performed using the ground truth bits as the labels. In particular, denoting the set of trainable parameters by $\boldsymbol{\theta}$, the loss function is defined as \cite{Honkala21}
\begin{align}
  {\mathrm{CE}}(\boldsymbol{\theta})\triangleq
  - \frac{1}{\#\mathcal DB}\sum_{(i,j)\in \mathcal D}\sum_{l=0}^{B-1} & \left(b_{ijl} \log(\hat{b}_{ijl}) \right.\nonumber\\
	&\left.+ (1 - b_{ijl}) \log(1 - \hat{b}_{ijl})\right)
\end{align}
where $\mathcal D$ denotes the time and frequency indices of data-carrying REs, $\#\mathcal D$ is the total number of data-carrying REs, and $\hat{b}_{ijl}$ is the receiver's estimate for the probability that the bit $b_{ijl}$ is one. The bit estimate is obtained by feeding the corresponding LLR through the sigmoid-function as
\begin{align}
  \hat{b}_{ijl} = \operatorname{sigmoid}\left(L_{ijl}\right) = \frac{1}{1 + e^{-L_{ijl}}},
\end{align}
where $L_{ijl}$ denotes the LLRs which are the actual output of the HybridDeepRx. The chosen stochastic gradient descent (SGD) algorithm in this work is the widely used Adam optimizer, which is updating the weights based on the CE loss in (3).

\section{Experimental Results}
\label{sec:exp}

\subsection{Data Generation}

\begin{table}[t]
  \setlength{\tabcolsep}{2pt}
    \renewcommand{\arraystretch}{1.3}
    \footnotesize
    \centering
    \vspace*{2mm}
    \caption{\textsc{Simulation parameters for training and validation}}
    \begin{tabular}{|l|l|l|}
    \hline
    \textbf{Parameter} & \textbf{Value}  & \textbf{Randomization}\\
		\hline\hline
		Channel model & AWGN & Noise realizations\\
		\hline
		PA model & Measured & Dithered coefficients\\
		\hline
		SNR & $0$ dB -- $30$ dB & Uniform distribution\\
		\hline
		Channel bandwidth & 5~MHz & None\\
		\hline
		Number of subcarriers ($N_D$) & 312 subcarriers & None\\
		\hline
		FFT size ($N$) & 512 & None\\
		\hline
		Subcarrier spacing & 15 kHz & None\\
		\hline
		Maximum CP length ($N_{CP}$) & 40 & None\\
		\hline
		OFDM symbol duration & 71.4 $\mu$s & None\\
		\hline
		TTI length ($N_\mathrm{symb}$) & 14 OFDM symbols & None\\
		\hline
		Modulation scheme & 16-QAM, 64-QAM & None\\
		\hline
    \end{tabular}
    \label{table:param}
\end{table}
  
In order to generate training data, we simulated a 5G physical uplink shared channel (PUSCH) link with Matlab's 5G Toolbox \cite{Matlab5G}, using the parameters specified in Table~\ref{table:param}. Moreover, to simulate the nonlinear behaviour of the PA, the response of a real-life PA module was measured under a high input power. Then, a 17th order polynomial was fitted to the measurements, representing the AM-AM and AM-PM response of the PA. To ensure that HybridDeepRx does not simply memorize the PA response, we introduce a dithering term that is applied to the measured PA polynomial to produce several slightly different PA models for training and validation, as also the true PA realizations vary across the UEs in real networks. The dithering is performed by adding a normally distributed random number to each polynomial coefficient, with a weight factor that is proportional to the magnitude of the original polynomial coefficient, while also imposing an applicable saturation level to the model such that physical PA behavior is correctly mimicked. Moreover, it should be noted that a different set of random PA polynomials are used for training and validation.

The datasets for training employ 30 dithered PA models with randomly chosen SNR in the range of 0 to 30 dB, the total size of the training dataset being 30~000 TTIs. Validation datasets employ 10 different dithered PA models and SNRs in uniform grid in the same range, containing 15~500 TTIs in total. Moreover, these datasets were generated separately for the considered 16-QAM and 64-QAM modulation schemes.

\subsection{Performance Evaluation}

The performance of the proposed network is first evaluated under varying levels of nonlinearity and an additive white Gaussian noise (AWGN) channel. We consider uncoded bit error rate (BER) as the main performance criteria. The results of the HybridDeepRx are compared with two alternative receivers: (\emph{i}) LMMSE with known channel, which in the AWGN case consists only of the phase and amplitude response of the PA, and (\emph{ii}) DeepRx as presented in \cite{Honkala21}. Moreover, also the theoretical AWGN BER is shown for reference. The comparison with DeepRx shows the impact of the time-domain pre-FFT ResNet blocks, while the AWGN BER provides the upper bound for the performance.

\begin{figure*}[ht]
    \centering
    \begin{subfigure}[t]{0.48\linewidth}
        \centering
        \includegraphics[width=\linewidth,trim={1cm 7.8cm 2cm 8cm},clip]{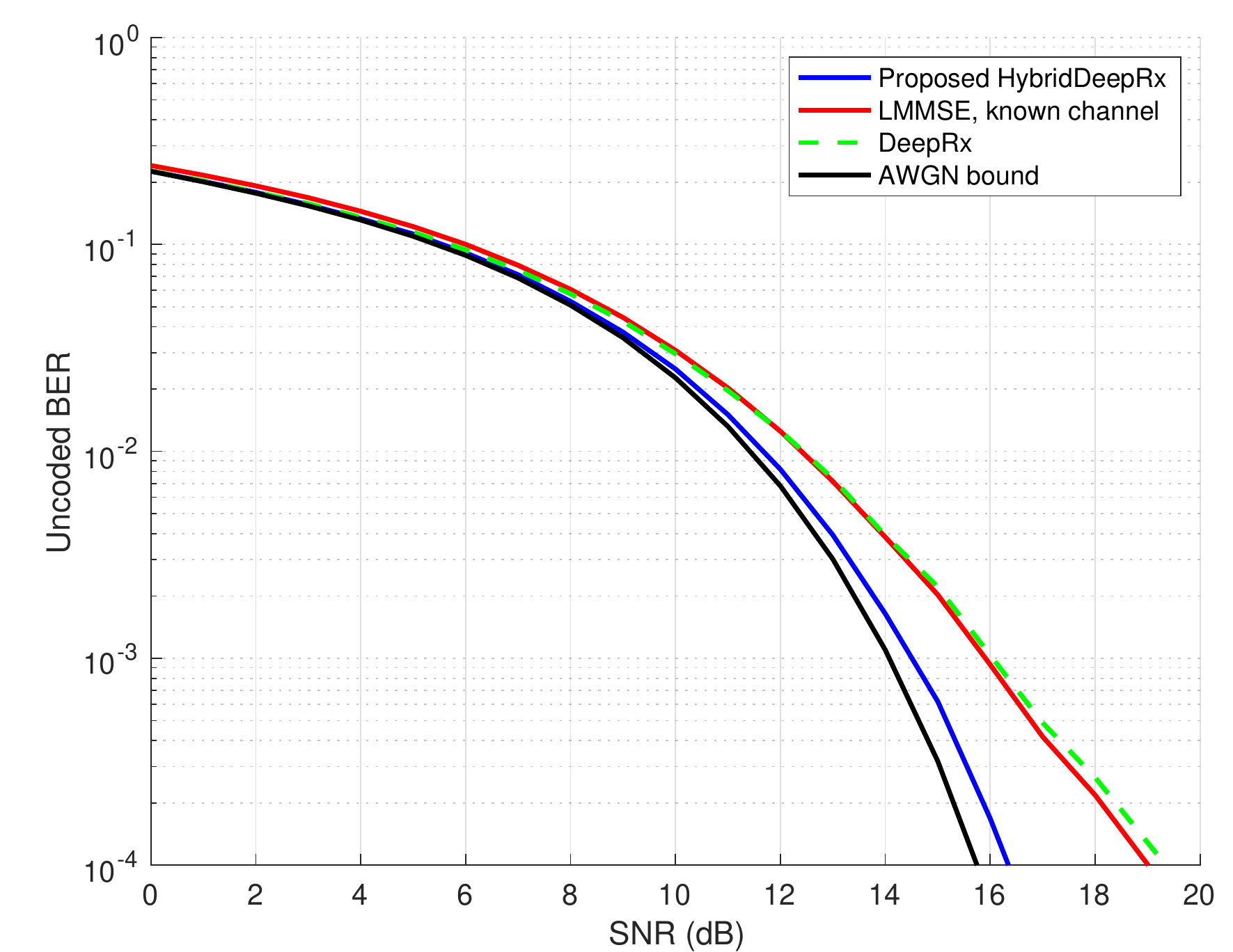}
        \caption{16-QAM}
        \label{fig:16QAM}
    \end{subfigure}\hfill
    \begin{subfigure}[t]{0.48\linewidth}
        \centering
        \includegraphics[width=\linewidth,trim={1cm 7.8cm 2cm 8cm},clip]{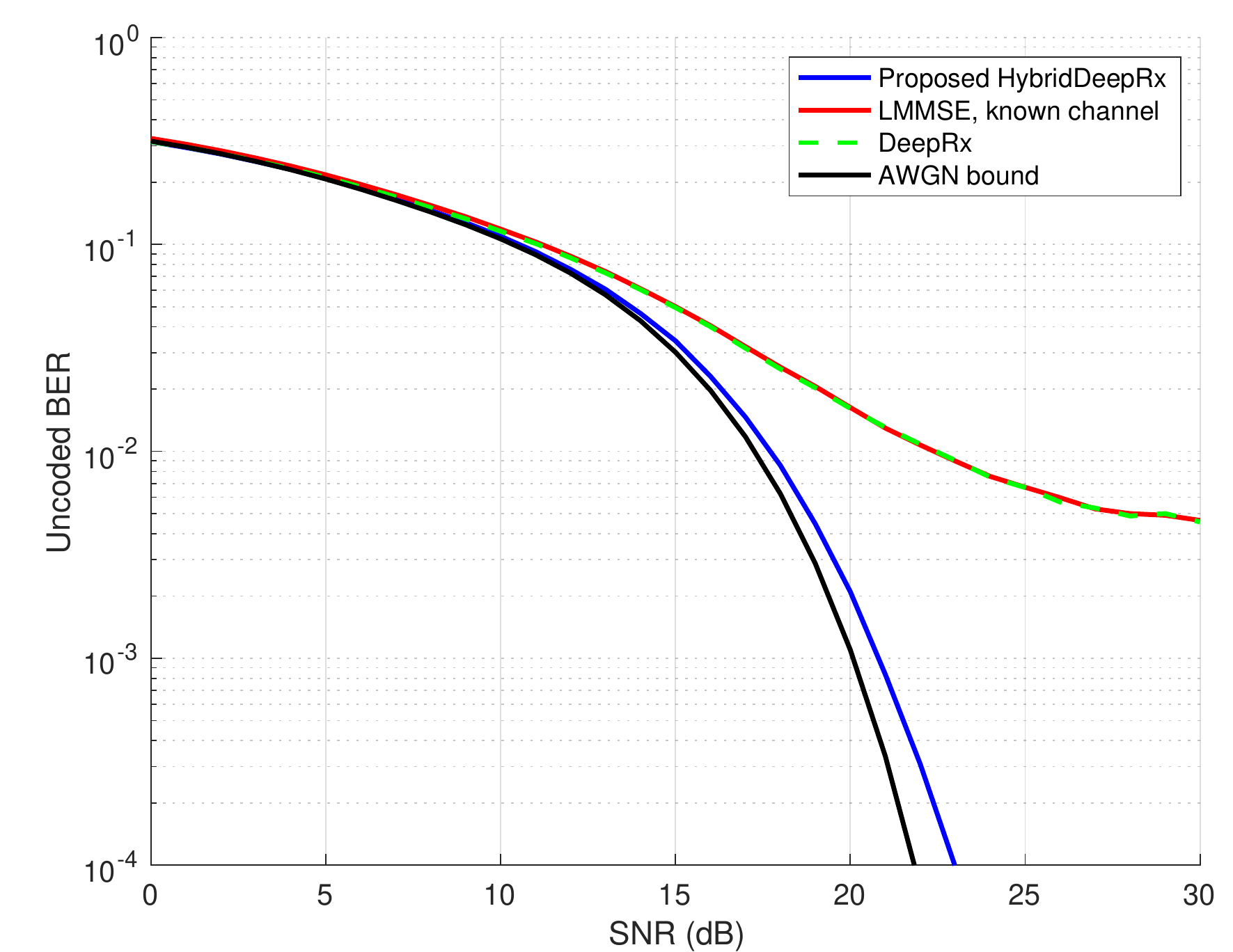}
        \caption{64-QAM}
        \label{fig:64QAM}
    \end{subfigure}
\caption{Uncoded BER performance of the proposed HybridDeepRx in comparison with prior art and benchmarks, under a PA backoff of 3~dB with (a) 16-QAM and (b) 64-QAM modulation schemes.}
\label{fig:64and16QAM}
\end{figure*}

Figure \ref{fig:64and16QAM} shows the BER performance when we set the PA backoff value to 3 dB, which corresponds to an EVM of roughly 8\%. This is the highest allowed EVM value for 64-QAM modulation in 3GPP 5G NR specifications. It is evident that with both considered modulation orders, HybridDeepRx has considerably better performance than the LMMSE or DeepRx benchmark receivers. In fact, HybridDeepRx almost achieves the AWGN bound. This clearly highlights the benefit of the temporal processing, achieved by the trained layers before the FFT.

\begin{table}[t] 
\setlength{\tabcolsep}{0.5pt}
    \renewcommand{\arraystretch}{1.3}
    \footnotesize
    \centering
    \vspace*{2mm}
    \caption{\textsc{Example link budget for illustrating uplink coverage extension at 3.5~GHz}} %
\begin{tabular}{|>{\raggedright}p{3.5cm}|>{\centering}p{2.2cm}|>{\centering\arraybackslash}p{2.2cm}|}
\hline
\textbf{Parameter}     & \textbf{LMMSE} & \textbf{HybridDeepRx} \\ \hline \hline
PA output backoff      & 4 dB       & 1 dB              \\ \hline
PA output power      & 26 dBm       & 29 dBm              \\ \hline
UE coupling losses     & \multicolumn{2}{c|}{4 dB}     \\ \hline
UE antenna gain        & \multicolumn{2}{c|}{0 dB}     \\ \hline
EIRP                   & 22 dBm     & 25 dBm            \\ \hline \hline
Noise power            & \multicolumn{2}{c|}{-107 dBm} \\ \hline
BS noise figure        & \multicolumn{2}{c|}{2 dB}     \\ \hline
SNR requirement        & \multicolumn{2}{c|}{19 dB}    \\ \hline
RX Sensitivity            & \multicolumn{2}{c|}{-83 dBm}  \\ \hline
BS coupling losses     & \multicolumn{2}{c|}{3 dB}     \\ \hline
BS antenna gain        & \multicolumn{2}{c|}{20 dB}    \\ \hline \hline
Maximum path loss      & 125 dB     & 128 dB            \\ \hline
Maximum distance, LOS  & 4731 m     & 5623 m (+19\%)    \\ \hline
Maximum distance, NLOS & 723 m      & 865 m (+19\%)     \\ \hline
\end{tabular}
\label{table:linkbudget}
\end{table}

\begin{figure*}[t!]
    \centering
    \begin{subfigure}[t]{0.48\linewidth}
        \centering
        \includegraphics[width=\linewidth,trim={1cm 7.6cm 2cm 8cm},clip]{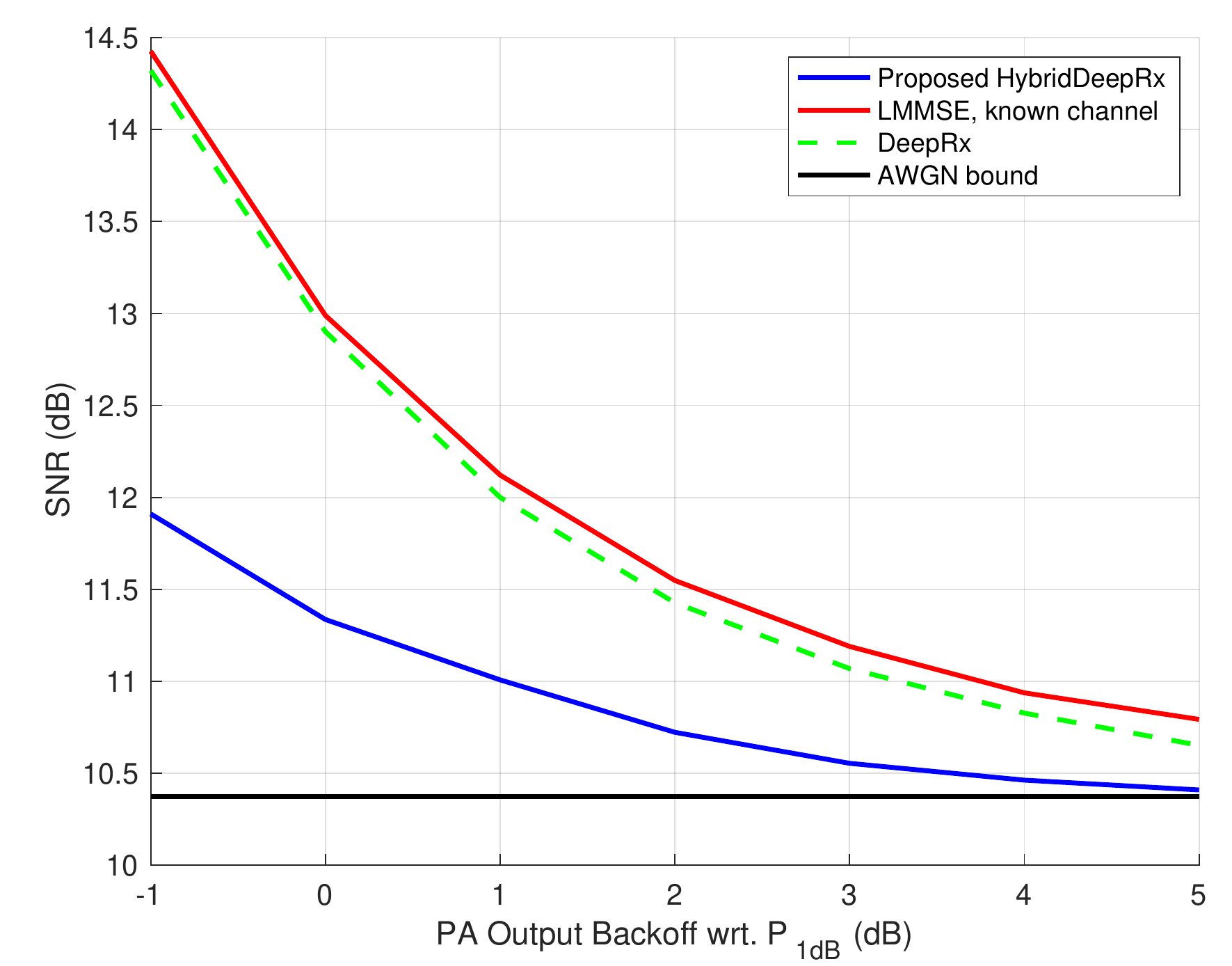}
        \caption{BER 10\%}
        \label{fig:BER10}
    \end{subfigure}\hfill
    ~
    \begin{subfigure}[t]{0.48\linewidth}
        \centering
        \includegraphics[width=\linewidth,trim={1cm 7.6cm 2cm 8cm},clip]{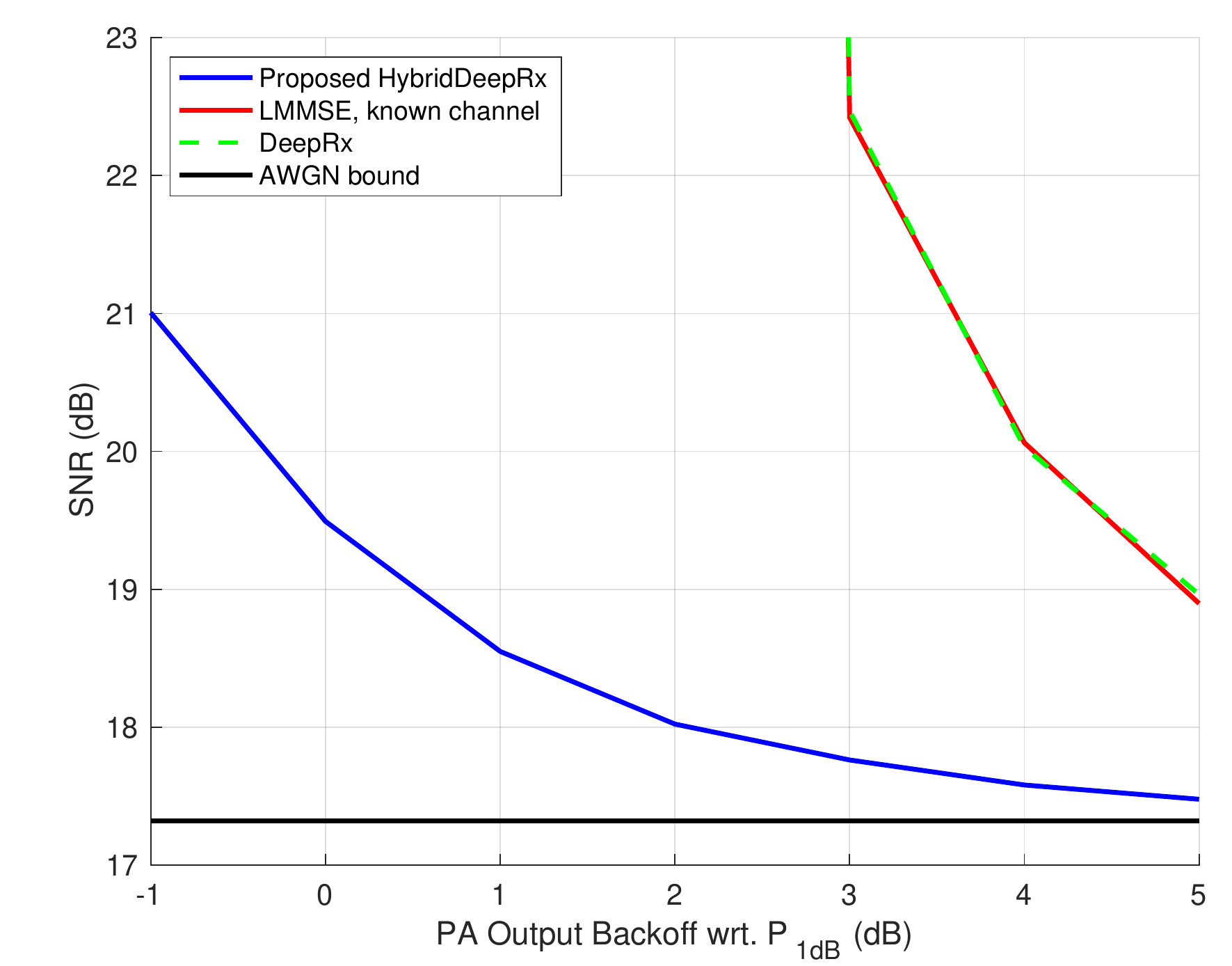}
        \caption{BER 1\%}
        \label{fig:BER1}
    \end{subfigure}
\caption{Performance of the proposed HybridDeepRx under varying levels of nonlinearity at (a) 10\% uncoded BER and (b) 1\% uncoded BER for 64-QAM.}
\label{fig:Backoff_BER}
\end{figure*}

Let us next investigate the performance with the higher-order 64-QAM modulation, considering a specific BER value. To this end, Fig.~\ref{fig:Backoff_BER} shows the SNR required to achieve uncoded BER values of 10\% and 1\% with respect to different levels of nonlinear distortion. Lower PA backoff value indicates higher nonlinearity. It can be observed that the proposed HybridDeepRx receiver can achieve the target BER with considerably lower SNR than the benchmark receivers. In fact, in Fig.~\ref{fig:BER1}, DeepRx and LMMSE receivers saturate before reaching 1\% BER within the studied SNR range if the PA backoff is less than 3~dB. As opposed to this, HybridDeepRx can achieve the BER target even under the most severe nonlinear distortion considered in this work. This indicates that the amount of nonlinear distortion is not a significant bottleneck for the detection performance of HybridDeepRx, thus allowing to push the transmitter PA system towards saturation for improved power-efficiency and coverage. 

\begin{figure}[t!]
	\centering
	\includegraphics[width=\columnwidth,trim={1cm 7.8cm 2cm 8cm},clip]{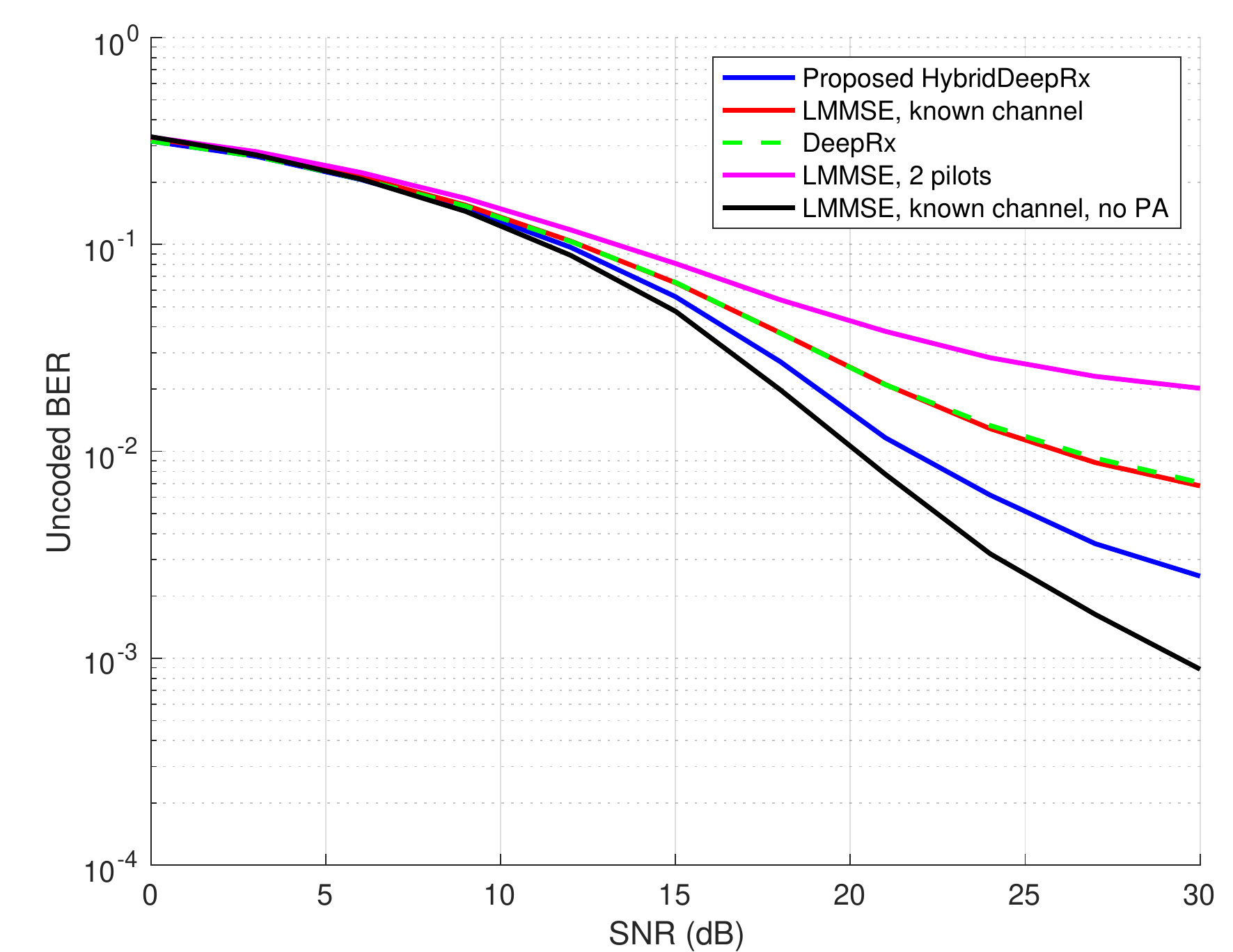}
	\caption{Uncoded BER performance of the proposed HybridDeepRx in a multipath channel with PA backoff 3 dB and 64-QAM. The channel has a maximum Doppler shift of 40 Hz and a delay spread 100 ns.}
	\label{fig:Multipath}
\end{figure}

We have also conducted preliminary analysis with multipath channels, utilizing the 3GPP tapped delay line (TDL) channel models \cite{NR_38901}. Figure~\ref{fig:Multipath} shows again the uncoded BER performance of the different receiver solutions. In addition to DeepRx and LMMSE with known channel, the BER is also shown for a practical LMMSE receiver estimating the channel from the DMRS symbols, and for LMMSE with known channel, but without the nonlinear PA. The last effectively represents an upper bound for the achievable performance. The results are largely in line with the AWGN scenario, the HybridDeepRx clearly outperforming the other solutions. Interestingly, the uncoded BER of HybridDeepRx is only 1--1.5~dB behind the case with a fully linear transmitter and known channel, which illustrates the high accuracy with which it can detect the distorted RX waveform.

Finally, to provide some further insight into the benefits of the proposed HybridDeepRx receiver technology, in Table \ref{table:linkbudget} we show an example link budget for the 5G NR uplink to illustrate the potential coverage extension enabled by the reduced PA backoff. In the given link budget, we have assumed a 5~MHz channel bandwidth at 3.5~GHz carrier frequency with a rural macro (RMa) path loss model (with default parameters) considering both line-of-sight (LOS) and non-line-of-sight (NLOS) conditions, as defined in \cite{NR_38901}. Moreover, the considered SNR requirement is chosen based on Fig.~\ref{fig:BER1}, where the SNR of 19~dB indicates the required non-negative backoffs for both the HybridDeepRx and LMMSE receivers to reach 1\% BER. By taking into account the potential nonlinear power increase at the PA output, the used PA output backoffs in the link budget are approximated as 1~dB and 4~dB for the HybridDeepRx receiver and the LMMSE receiver, respectively. As shown in Table~\ref{table:linkbudget}, the proposed HybridDeepRx receiver algorithm increases the link coverage by 19\% compared to the baseline LMMSE receiver. For example, in case of NLOS propagation, this indicates an absolute coverage extension from 723~m to 865~m in link distance.

\vspace{-0mm}
\section{Conclusions}
\label{sec:conc}
In this paper, we presented a novel deep learning based receiver solution, the so-called HybridDeepRx, tailored for accurate detection of nonlinearly distorted signals with high EVM. This is achieved by introducing trainable convolutional layers both in time and frequency domains, where the former is particularly suited for handling the nonlinear distortion, while the latter performs the actual signal detection. The proposed receiver architecture is shown to be able to detect even heavily distorted signals with a considerably high EVM, while the benchmark receivers fail to detect such signals reliably. Indeed, the performance gain compared to a conventional linear receiver is several dBs even with reasonable levels of nonlinear distortion. The proposed HybridDeepRx also outperforms in the high-EVM scenario the previously presented ML-based DeepRx receiver which utilizes convolutional layers only in frequency domain. These findings pave the way towards more power efficient radios where the effects of hardware impairments can be handled with the help of deep learning aided receiver solutions. Our future work will include development of other related disruptive receiver schemes, such as involving convolutional layers in time domain while having more ordinary LMMSE receiver in frequency domain, as well as devising corresponding training solutions for such concepts.

\section*{Acknowledgment}
This work was supported in part by Business Finland under the project 5G VIIMA, and in part by Academy of Finland under the grants \#319994 and \#332361.

\bibliographystyle{IEEEtran}
\bibliography{Bibliography}

\end{document}